\documentclass[sigconf, nonacm]{acmart}

\newcommand\vldbdoi{XX.XX/XXX.XX}

\newcommand\vldbvolume{18}
\newcommand\vldbissue{X}

\newcommand\vldbavailabilityurl{URL_TO_YOUR_ARTIFACTS}
\newcommand\vldbpagestyle{plain} 

\usepackage{enumitem}
\usepackage{environ}
\usepackage{tabularx}

\usepackage{cleveref}
\crefformat{chapter}{\S#2#1#3}
\crefformat{section}{\S#2#1#3}
\crefformat{subsection}{\S#2#1#3}
\crefformat{subsubsection}{\S#2#1#3}

\usepackage{xspace}
\usepackage{soul}
\usepackage{caption}
\usepackage{subcaption}
\usepackage{booktabs}
\usepackage{multirow}
\usepackage{fancyvrb}
\usepackage{algorithm}
\usepackage[noEnd=true,indLines=true,beginComment=//~,beginLComment=//~,endLComment=~]{algpseudocodex}

\usepackage{balance}
\usepackage{pifont}

\usepackage{cuted}
\newcommand{\SYS}{\textsc{Keigo}\xspace}

\definecolor{navy}{HTML}{2F729C}
\definecolor{darkgray}{HTML}{323232}
\definecolor{seagreen}{HTML}{228B22}

\newcommand{\repository}{\url{https://github.com/dsrhaslab/keigo}}

\newcommand{\pmem}{NVMM\xspace}
\newcommand{\nvme}{NVMe SSD\xspace}
\newcommand{\sata}{SATA SSD\xspace}

\newcommand{\addnoteauthor}[2]{%
  \expandafter\newcommand\csname #1\endcsname[1]{%
    {%
      \small\color{black}%
      \fbox{\scriptsize\bfseries\sffamily#1}%
      $\blacktriangleright${\sffamily\itshape\color{#2}##1}$\blacktriangleleft$%
      }%
    }%
  \NewEnviron{#1block}{%
    \vspace*{.4\baselineskip}%
    \small\color{black}%
    \setlength\parskip{0pt plus 0pt minus 0pt}%
    \par\noindent\fbox{\scriptsize\bfseries\sffamily#1}%
    \par\nopagebreak\noindent\sffamily\itshape\color{#2}%
    \begin{tabularx}{\linewidth}{|X|}%
      \hline%
      \vspace{-2pt}%
      \BODY%
      \par \kern-6pt%
      \tabularnewline \hline%
    \end{tabularx}%
    \vspace*{.4\baselineskip}%
    }%
  \NewEnviron{#1itemize}{%
    \vspace*{.4\baselineskip}%
    \small\color{black}%
    \setlength\parskip{0pt plus 0pt minus 0pt}%
    \par\noindent\fbox{\scriptsize\bfseries\sffamily#1}%
    \par\nopagebreak\noindent\sffamily\itshape\color{#2}%
    \setlist{leftmargin=*,topsep=0pt,nosep}%
    \begin{tabularx}{\linewidth}{|X|}%
      \hline%
      \vspace{-2pt}%
      \begin{itemize}%
        \BODY%
      \end{itemize}%
      \par \kern-8pt%
      \tabularnewline \hline%
    \end{tabularx}%
    \vspace*{.4\baselineskip}%
    }%
  }

\addnoteauthor{rgmacedo}{orange}
\addnoteauthor{jtpaulo}{green}
\addnoteauthor{oana}{red}
\addnoteauthor{ruben}{teal}
\addnoteauthor{joe}{grey}
\addnoteauthor{zhongjie}{purple}

\newcommand{\revision}[1]{#1}

\renewcommand{\paragraph}[1]{\vspace{.2\baselineskip}\noindent\textbf{#1.}}
\newcommand{\emphparagraph}[1]{\vspace{.2\baselineskip}\noindent\emph{#1.}}
\newcommand{\emptyparagraph}{\vspace{.2\baselineskip}\noindent}

\hypersetup{
    colorlinks=true,
    linkcolor=navy,
    urlcolor=navy,
}

\usepackage{thmtools}
\usepackage[framemethod=TikZ]{mdframed}
\mdfsetup{skipabove=1em,skipbelow=0em, innertopmargin=5pt, innerbottommargin=6pt}
\theoremstyle{definition}

\definecolor{DimGray}{rgb}{0.41, 0.41, 0.41}
\colorlet{thmgreencolor}{DimGray!70!black}

\makeatletter

\declaretheoremstyle[
  headfont=\bfseries, bodyfont=\normalfont\itshape,
  mdframed={
    innertopmargin=2pt,
    innerleftmargin=8pt,
    innerrightmargin=8pt,
    innerbottommargin=5pt,
    linewidth=2pt,
    rightline=false, topline=false, bottomline=false,
    linecolor=black, backgroundcolor=DimGray!5,
  }
  ]{thmgraybox}
  
\makeatother

\declaretheorem[style=thmgraybox, numbered=yes, name=Takeaway]{takeaway}

\AtBeginEnvironment{takeaway}{
  \captionsetup{labelfont={color=thmgraycolor}}
}

\definecolor{Gray}{gray}{0.9}
\newcolumntype{?}{!{\vrule width 1.5pt}}

\algnewcommand{\InlineIf}[2]{
  \State \algorithmicif\ #1\ \algorithmicthen\ #2}

\begin{document}


\title{\SYS: Co-designing Log-Structured Merge Key-Value Stores with a Non-Volatile, Concurrency-aware Storage Hierarchy\\(Extended Version)}

\author{R{\'u}ben Ad{\~a}o}
\affiliation{%
  \institution{INESC TEC \& University of Minho}
}
\email{ruben.d.adao@inesctec.pt}

\author{Zhongjie Wu}
\affiliation{%
  \institution{Yale University}
}
\email{zhongjie.wu@yale.edu}

\author{Changjun Zhou}
\affiliation{%
  \institution{McGill University}
}
\email{changjun.zhou2@mail.mcgill.ca}

\author{Oana Balmau}
\affiliation{%
  \institution{McGill University}
}
\email{oana.balmau@mcgill.ca}

\author{Jo{\~a}o Paulo}
\affiliation{%
  \institution{INESC TEC \& University of Minho}
}
\email{joao.t.paulo@inesctec.pt}

\author{Ricardo Macedo}
\affiliation{%
  \institution{INESC TEC \& University of Minho}
}
\email{ricardo.g.macedo@inesctec.pt}

\begin{abstract}
    We present \SYS, a concurrency- and workload-aware storage middleware that enhances the performance of log-structured merge key-value stores (LSM KVS) when they are deployed on a hierarchy of storage devices. 
    The key observation behind \SYS is that there is no \emph{one-size-fits-all} placement of data across the storage hierarchy that optimizes for all workloads. Hence, to leverage the benefits of combining different storage devices, \SYS places files across different devices based on their parallelism, I/O bandwidth, and capacity.
    We introduce three techniques -- \emph{concurrency-aware data placement}, \emph{persistent read-only caching}, and \emph{context-based I/O differentiation}. 
    \SYS's is portable across different LSMs, is adaptable to dynamic workloads, and does not require extensive profiling. Our system enables established production KVS such as RocksDB, LevelDB, and Speedb to benefit from heterogeneous storage setups.


    

    We evaluate \SYS using synthetic and realistic workloads, showing that it improves the throughput of production-grade LSMs up to 4$\times$ for write- and 18$\times$ for read-heavy workloads when compared to general-purpose storage systems and specialized LSM KVS.

    \vspace*{-20pt}
\end{abstract}

\maketitle

\pagenumbering{arabic} 

\pagestyle{\vldbpagestyle}
\begingroup
\renewcommand\thefootnote{}\footnote{\noindent
This work is licensed under the Creative Commons BY-NC-ND 4.0 International License. Visit \url{https://creativecommons.org/licenses/by-nc-nd/4.0/} to view a copy of this license. For any use beyond those covered by this license, obtain permission by emailing \href{mailto:info@vldb.org}{info@vldb.org}. Copyright is held by the owner/author(s). Publication rights licensed to the VLDB Endowment. \\
\raggedright Proceedings of the VLDB Endowment, Vol. \vldbvolume, No. \vldbissue\ %
ISSN 2150-8097. \\
\href{https://doi.org/\vldbdoi}{doi:\vldbdoi} \\
}\addtocounter{footnote}{-1}\endgroup

\ifdefempty{\vldbavailabilityurl}{}{
\vspace{.3cm}
\begingroup\small\noindent\raggedright\textbf{PVLDB Artifact Availability:}\\
The source code, data, and/or other artifacts have been made available at \repository.
\endgroup
}

\renewcommand{\thefigure}{\arabic{figure}}
\setcounter{figure}{0}

\textbf{}\section{Introduction}
\label{sec:introduction}

Log-structured merge-tree (LSM) key-values stores (KVS), such as RocksDB~\cite{rocksdb-git}, LevelDB~\cite{leveldb-git}, and Speedb~\cite{speedb-git}, have become a fundamental storage building block for a variety of data-intensive applications, including databases~\cite{MyRocks:2020:Matsunobu,cockroachdb,mongodb,foundationdb}, caching systems~\cite{Kangaroo:2022:Mcallister,CacheLib:2020:Berg}, file systems~\cite{alluxio,Ceph:2019:Aghayev}, and analytics engines~\cite{DataProcessingFacebook:2016:Chen,DragonFacebook,LSM:2020:Luo}.
Their wide adoption is driven by their natural fit for write-heavy workloads. 
They buffer writes in main memory, flush them as sorted files to storage, and merge (compact) files across multiple levels of increasing capacities~\cite{LSM:2020:Luo,Log:1996:Oneil}.
However, the amount of data stored in these KVS is growing exponentially (in the order of hundreds of GiBs to TiBs), raising performance concerns as many applications require high throughput and low tail latency~\cite{SILK:2019:Balmau,Kvell:2019:Lepers}. 

To overcome this challenge, prior work proposes heterogeneous storage hierarchies combining the \emph{performance} of emerging non-volatile byte-addressable technologies (\emph{e.g.,} 3D XPoint~\cite{3DXPoint:2017:Hady}, PCM \cite{PCM:2022:Aljameh}) with the \emph{storage capacity} of traditional block-addressable devices (\emph{e.g.,} \nvme, \sata). 
These solutions are provided as general-purpose hierarchical storage systems~\cite{OpenCAS,P2Cache:2023:Lin,Strata:2017:Kwon,Ziggurat:2019:Zheng,SPFS:2023:Woo}, serving as back-end to different applications (including LSM KVS), or as specialized KVS purposely built for heterogeneous storage~\cite{PrismDB:2023:Raina,BushStore:2024:Wang,MatrixKV:2020:Yao,SpanDB:2021:Chen}. 
In this paper, we classify storage devices into \emph{performance devices} and \emph{capacity devices}.
\emph{Performance devices} offer high performance but reduced storage space (\emph{e.g.,} \pmem), while \emph{capacity devices} offer larger and cheaper storage alternative (\emph{e.g.,} \nvme, \sata).
Existing systems typically prioritize placing as much of the LSM structure as possible in the \emph{performance device}, starting with the performance-sensitive components of the LSM, \emph{i.e.,} the commit log (C$_{log}$) and top levels of the LSM tree (L$_0$ and L$_1$), followed by filling any leftover space with data from other levels.
The remaining levels of the LSM are placed on the \emph{capacity device}.

However, we show that \emph{placing LSM components across complex and heterogeneous storage hierarchies solely based on the I/O bandwidth and capacity of devices leads to degraded performance}.
To understand how the use of different storage devices impacts the performance of LSM-based KVS, and as our first contribution, we conduct an experimental study combining \pmem, \nvme, and \sata devices, where we report the following key findings (\cref{sec:study}). 
First, under write-heavy workloads, placing as many LSM components as possible on the \emph{performance device} is actually detrimental, degrading throughput up to 35\% when compared to placing a smaller subset of data items. 
This phenomenon is caused by the increased write concurrency placed on the \emph{performance device}, stemming from foreground writes to C$_{log}$, flushing in-memory data to L$_0$, and multiple parallel compactions occurring over different levels of the LSM tree.
Second, since storage devices scale better for concurrent reads, placing as many components in the \emph{performance device} is beneficial for read-intensive workloads. 
This means that there is no winning placement strategy that simultaneously fits the requirements of read and write workloads.

Motivated by these findings, as our second contribution, we propose \SYS, a concurrency- and workload-aware storage middleware that accelerates the performance of LSM-based KVS for both read- and write-intensive workloads.
The novelty of \SYS lies in how it carefully places each KVS component across the hierarchy of storage devices by consolidating the inherent properties of LSMs with the characteristics of each device, including \emph{device-level parallelism}, \emph{I/O bandwidth}, and \emph{storage capacity}. 
\SYS can be used with multiple LSM KVS to provide a better placement of LSM components over the storage hierarchy. 
It is adaptable to changing read:write ratios and skew, and only needs minimal device profiling to create an initial data placement scheme.

\SYS introduces three techniques.
To optimize write workflows, the LSM levels placed in the \emph{performance device} are not only determined by the device I/O bandwidth and capacity but also by the number of concurrent writers that can work in parallel (\cref{subsec:optimizing-writes}). 
This means that under write-heavy workloads, \SYS stores just a small subset of files of the KVS (\emph{i.e.,} C$_{log}$ and lower levels of the LSM tree) on the faster device, reducing write contention. 
Second, for read-heavy workloads, \SYS introduces a per-device persistent read-only cache that continuously tracks and copies hot KVS files from slower to faster storage devices, maximizing the hit ratio on the upper levels of the hierarchy (\cref{subsec:optimizing-reads}).
Third, for easy compatibility with different LSM KVS, we use context propagation techniques~\cite{ContextPropagation:2018:Mace,PAIO:2022:Macedo}. 
We expose a POSIX-like interface where system calls propagate the internal KVS operation that started a given POSIX request (\emph{e.g.,} C$_{log}$, flush, compactions) to \SYS, which uses this information to make efficient placement decisions (\cref{subsec:context-propagation}).


As our third contribution, we validate the performance of \SYS through a comprehensive experimental evaluation, using both synthetic (\emph{i.e.,} YCSB benchmark~\cite{YCSB:2010:Cooper}) and production workloads from Meta~\cite{FacebookTraces:2020:Cao} and Nutanix~\cite{Kvell:2019:Lepers}. 
We demonstrate that \SYS can enhance a range of LSM KVS by adding it to three popular systems with less than 100~LoC: RocksDB, LevelDB, and Speedb. 
Further, we show that \SYS improves LSM KVS performance in two storage hierarchies, composed of DRAM-\pmem-NVMe and DRAM-NVMe-{\sata}s.
We compare \SYS against general-purpose storage systems, namely ext4 and OpenCAS~\cite{OpenCAS}, and state-of-the-art LSMs designed for heterogeneous storage devices, namely BushStore~\cite{BushStore:2024:Wang} and PrismDB~\cite{PrismDB:2023:Raina}.
Results show that \SYS outperforms all systems across all testing scenarios.
Compared to general-purpose storage systems, \SYS improves RocksDB throughput up to 4$\times$ and 12$\times$ under write-heavy and read-heavy workloads, respectively.
As for the specialized LSM-based KVS, \SYS improves RocksDB throughput up to 1.6$\times$ for write-heavy workloads and up to 18$\times$ for read-heavy workloads.
Moreover, for Meta and Nutanix production workloads, \SYS improves the next best system by 1.15$\times$ and 1.4$\times$, respectively.
Finally, \SYS enhances the performance of Speedb and LevelDB's baselines by up to 8.85$\times$ and 5.76$\times$.

\section{Background}
\label{sec:background}

This section provides background on devices that make up current storage hierarchies and discusses classic strategies to manage them. 
Further, it provides background on the organization of LSM KVS.


\subsection{Heterogeneous storage management}
\label{subsec:heterogeneous-management}

Data centers use heterogeneous storage hierarchies composed of different storage devices, each offering trade-offs of performance, capacity, and cost.
Emerging \pmem devices (\emph{e.g.,} 3D XPoint~\cite{3DXPoint:2017:Hady}, PCM \cite{PCM:2022:Aljameh}) enable byte-addressable persistent storage with performance closer to that of DRAM. 
Ultra-low latency {\nvme}s (\emph{e.g.,} Z-NAND~\cite{ZNAND:2018:Cheong}) deliver $\mu$s-scale latencies with larger capacity than \pmem, while traditional block-addressable devices (\emph{e.g.,} \sata, HDD) provide a denser and cheaper alternative.

To manage a heterogeneous storage hierarchy, systems mainly follow two strategies: \emph{caching} and \emph{tiering}.
For simplicity, we assume a two-tier hierarchy, made of a \emph{performance device} that is faster, smaller, and expensive (\pmem), and a \emph{capacity device} that provides cheaper and large capacity storage (\nvme). 
Our design, however, can be used for multi-tiered storage hierarchies (\cref{sec:design}).


\paragraph{Caching}
In caching, the \emph{performance device} is used as a persistent cache to accelerate the \emph{capacity device}~\cite{SPFS:2023:Woo,OpenCAS,P2Cache:2023:Lin}.
Hot data resides in the \emph{performance device}, ensuring low latency and high throughput under skewed read-heavy workloads.
On cache misses, the \emph{capacity device} serves the requests.
The item placement is determined by an eviction policy (\emph{e.g.,} LRU, LFU). 
Depending on the writing policy (\emph{e.g.,} write-back), the \emph{performance device} can absorb write operations, flushing dirty data upon explicit synchronization. 

\paragraph{Tiering}
In tiering, data items are partitioned across devices according to a specific placement scheme, which is often driven by the items' popularity, size, consistency guarantees, and more~\cite{Ziggurat:2019:Zheng,PrismDB:2023:Raina,Strata:2017:Kwon}.
Contrary to caching, items only reside in one of the devices and are not constantly promoted/evicted to/from the \emph{performance device}.


\subsection{Heterogeneous storage in LSM}
\label{subsec:heterogeneous-lsm}


\paragraph{LSM overview}
Log-structured merge tree (LSM) key-value stores (KVS), such as RocksDB~\cite{rocksdb-git}, LevelDB~\cite{leveldb-git}, and Cassandra~\cite{Cassandra:2010:Lakshman}, are widely adopted storage systems that are optimized for write-intensive workloads \cite{LSM:2020:Luo,Log:1996:Oneil}.
Write operations are absorbed by a memory component (C$_m$ or \emph{memtable}) that when is full, it is \emph{flushed} to persistent storage in one large sequential I/O operation.
The flushed C$_m$ is then merged by background threads in a tree-like structure maintained in persistent storage (C$_{disk}$).
C$_{disk}$ contains multiple levels of increasing sizes (\emph{L}$_0$, \emph{L}$_1$, ..., \emph{L}$_N$), where each level contains multiple \emph{immutable} sorted files, called SSTs.
This merging operation is called \emph{compaction}.
LSM KVS can run several concurrent compactions using dedicated background threads, in addition to the foreground load.
To avoid losing data held in C$_m$, writes are backed up in a \emph{write-ahead log} (C$_{log}$) that also resides in persistent storage.
To improve read performance, data items may be temporarily held in an in-memory region called \emph{block cache}.
Foreground reads first access C$_m$, followed by the block cache, followed by the SST files.

With this multi-level structure, LSMs are a natural fit for leveraging storage hierarchies. 
So far, heterogeneous storage in LSMs has been tackled through two main approaches.

\paragraph{Using general-purpose tiered storage systems}
A commonly used approach is to use general-purpose hierarchical storage systems (\emph{e.g.,} OpenCAS~\cite{OpenCAS}, Ziggurat~\cite{Ziggurat:2019:Zheng}, P2Cache~\cite{P2Cache:2023:Lin}), which usually reside at the kernel and do not require POSIX-compliant applications to be modified to obtain performance gains.
The storage system determines which device services each I/O request based on the data access pattern (\emph{e.g.,} \emph{write} followed by \emph{fsync}~\cite{P2Cache:2023:Lin,SPFS:2023:Woo}, small \emph{vs.} large writes~\cite{Ziggurat:2019:Zheng,TPFS:2023:Zheng}).
However, while these systems are designed to handle a wide range of applications, they are agnostic of LSM I/O logic, treating all requests in the same way regardless of their origin or priority.
For example, while compactions at different levels exhibit similar access patterns --- sequentially read SST files from disk, merge sort them in memory, and sequentially write new SST files to disk --- they incur different performance costs~\cite{SILK:2019:Balmau,ADOC:2023:Yu}.

\paragraph{Building new LSMs from the ground-up}
Alternatively, prior work proposes building new LSM KVS for heterogeneous storage.
These new designs consider the trade-offs of each storage device with careful placement of the LSM components~\cite{SpanDB:2021:Chen,WaLSM:2023:Chen,BushStore:2024:Wang}, employ new compaction schemes compliant with the storage hierarchy~\cite{PrismDB:2023:Raina,MatrixKV:2020:Yao}, and redesign the write operation flow~\cite{SpanDB:2021:Chen,NoveLSM:2018:Kannan}.
However, such systems are difficult to adopt in practice, as LSMs are typically core components in large pipelines. 
A significant implementation effort is required to replace existing LSMs with such alternatives.
This problem becomes further accentuated when new storage devices are released, deprecating the previous optimizations. 

\emptyparagraph
\SYS provides a middle-ground between these approaches. 
By offering a middleware tailored for LSM KVS, \SYS is aware of the I/O flows and priorities unique to LSMs, leveraging this information when deciding the placement of different components.
Moreover, its design is flexible: it can be used by various LSM KVS and can be adapted to heterogeneous storage hierarchies of different depths.

\section{Issues with heterogeneous storage for LSM}
\label{sec:study}

We conduct an experimental study to understand how hierarchies of different storage devices impact the performance of LSM KVS.
We consider hierarchies combining byte-addressable and block-based storage devices.
Our observations \emph{complement} previous studies that explore idiosyncrasies of individual storage devices~\cite{OptaneSSD:2019:Wu,UnwrittenContractSSD:2017:He,PMMStudy:2020:Yang,PmemIdiosyncrasies:2020:Gugnani,WritesHurt:2022:Fedorova}, by pinpointing their impact in LSM-based KVS.

\paragraph{Hardware and OS configurations}
We run the experiments in a server with two 18-core Intel Xeon processors, 192~GiB of memory, a 128~GiB Intel Optane DCPMM (AppDirect mode), \revision{a 1.6~TiB Intel P5600 NVMe SSD (PCIe 4.0 $\times$4)}, a 1.6~TiB Dell PM1725b NVMe SSD \revision{(PCIe 3.0 $\times$4)}, and a 480~GiB Intel S4610 SATA SSD, using Ubuntu Server 20.04 with kernel 5.4.0.
We restrict the main memory to 16~GiB using Linux control groups to ensure \emph{1)} the storage and memory configurations of the server form a hierarchy and \emph{2)} that most of the requests are submitted to persistent storage. 
\revision{Further, we use a single socket to prevent NUMA effects.
Unless stated otherwise, experiments containing an NVMe device were conducted with the PM1725b \nvme.}

\paragraph{LSM KVS configuration}
The experiments were conducted over RocksDB configured with two C$_m$ of 128~MiB, a thread pool of four threads for internal operations (including a flushing thread), and a 1~GiB block cache. 
These configurations correspond to production settings at Nutanix~\cite{SILK:2019:Balmau} and follow RocksDB tuning guidelines~\cite{rocksdb-tuning-guide}.
We evaluate RocksDB with 8 client threads using write-intensive (YCSB A, 1:1 r:w ratio)  and read-only (YCSB C) workloads with a uniform key distribution.
Before each experiment, RocksDB was pre-loaded with 50M key-value pairs (1KB value).

\paragraph{Storage setups}
The experiments were conducted over the following storage configurations:
\emph{1)} \emph{ext4} corresponds to an ext4 file system mounted over the NVMe or SATA SSDs;
\emph{2)} \emph{ext4-DAX} corresponds to \pmem backed by an ext4 file system with DAX enabled to perform direct I/O;
\emph{3)} we implemented a custom backend that combines \pmem (backed by a PMDK driver~\cite{pmdk}) and \nvme (ext4 file system). 
The first two setups show storage systems commonly used in production, while the latter demonstrates a storage hierarchy adopted by prior work~\cite{MatrixKV:2020:Yao,BushStore:2024:Wang,BushStore:2024:Wang,PrismDB:2023:Raina,SPFS:2023:Woo,P2Cache:2023:Lin}.


\subsection{Device-level parallelism}
\label{subsec:motivation-device-parallelism}

\begin{figure}[t]
    \centering
    \includegraphics[width=1\linewidth,keepaspectratio]{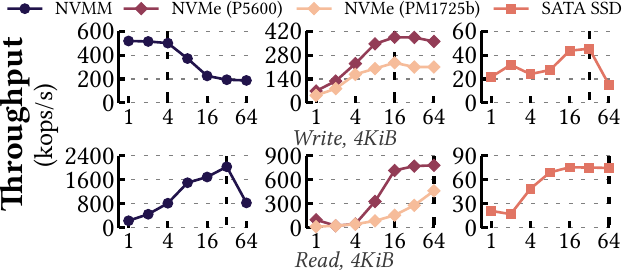}
    \caption{Performance of \pmem, two {\nvme}s, and \sata storage devices for FIO write and read workloads with increasing number of threads (1 to 64, x-axis). 
    Vertical lines mark the maximum performance achieved by each device.}
    \label{fig:motivation-fio}
    \vspace*{-10pt}
\end{figure}


Before exploring the performance of LSM under heterogeneous storage, we first show how each storage device performs when exposed to different workloads and concurrency levels.
We conducted experiments using the \texttt{fio} benchmark with sequential read and write workloads (\texttt{sync} I/O engine) under 4~KiB blocks with increasing worker threads (1 to 64).
Each worker operates over a distinct device region to avoid overlapping working sets.
We report the mean result over 5 runs, with less than 5\% standard deviation.

Figure~\ref{fig:motivation-fio} depicts the throughput of the \pmem, {\nvme}s, and \sata devices.
Storage devices have different performance and degrees of parallelism.
For \emph{write workloads}, \pmem reaches its peak performance under 4 threads (500~kops/s), while \nvme scales up to 16 threads.
\nvme devices are highly parallel, containing multiple I/O channels connected to independent flash dies.
While the \sata achieves the highest parallelism level, it experiences the lowest performance due to its limited interface bandwidth and architecture~\cite{UnwrittenContractSSD:2017:He}.
Beyond 4 workers, \pmem's performance decreases up to 2.6$\times$ (188~kops/s under 64 threads). 
As observed in~\cite{PMMStudy:2020:Yang,PmemIdiosyncrasies:2020:Gugnani}, the reason behind \pmem's poor scalability is caused by contention in the XPBuffer (\emph{i.e.,} increased evictions and write backs to the memory media) and the integrated memory controller (\emph{i.e.,} limited queue capacity when multiple cores target a single DIMM).
Consequently, \emph{under severe write contention, the {\nvme}s achieves similar or better performance than \pmem} --- under 16, 32, and 64 threads, the \pmem achieves 226~kops/s, 194~kops/s, and 188~kops/s, while \revision{the {\nvme}s achieve 235-384~kops/s, 210-383~kops/s, and 210-360~kops/s}, respectively.
For \emph{read workloads}, all storage devices showcase a higher parallelism level.
Specifically, \pmem achieves its maximum performance under 32 threads, while block-based storage devices can scale up to 64 threads.
Contrary to write workloads, \pmem read performance is limited by the number of physical NUMA cores~\cite{PMMStudy:2020:Yang}.

\revision{Moreover, despite offering the same degree of parallelism, {\nvme}s exhibit significantly different performance.
Specifically, the P5600 outperforms the 1725b up to 1.91$\times$ for write workloads and 4.65$\times$ for read workloads. 
This disparity stems from the bandwidth limits of their respective interfaces and interconnects, namely PCIe 4.0 for the P5600 device and PCie 3.0 for the PM1725b.}

\begin{takeaway} 
    Due to the characteristics of each storage device, device I/O bandwidth should not serve as the delimiting factor for managing a heterogeneous storage hierarchy.  
    Devices must be carefully combined to realize the full potential of the storage hierarchy in terms of performance, parallelism, and capacity.
\end{takeaway}


\subsection{Placement of LSM components}
\label{subsec:motivation-lsm-placement}

\begin{figure}[t]
    \centering
    \includegraphics[width=1\linewidth,keepaspectratio]{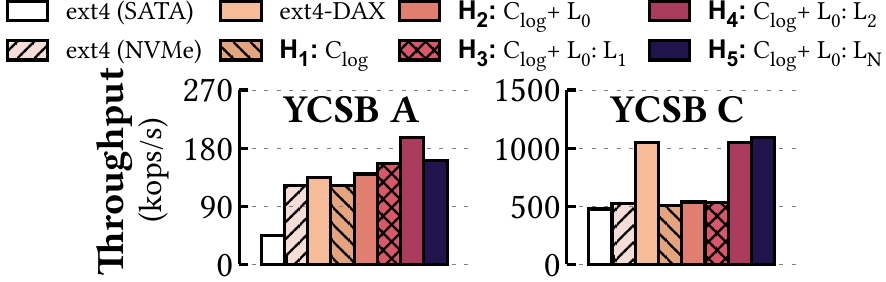}
    \caption{RocksDB performance when placing LSM components across a \pmem-\nvme hierarchy under mixed and read-only workloads.}
    \label{fig:motivation-lsm-components}
    \vspace*{-10pt}
\end{figure}

We now explore the impact of placing different LSM components (excluding C$_m$) across the storage hierarchy.
To this end, we compare RocksDB throughput under ext4 with NVMe and SATA SSDs, ext4-DAX, and the custom storage backend configured with different placement schemes.
In the latter, we start by placing C$_{log}$ on \pmem and keep the remainder components on the \nvme (H$_1$).
We then incrementally place more components on \pmem --- C$_{log}$+L$_0$ (H$_2$), C$_{log}$+L$_0$:L$_1$ (H$_3$), ... --- until the full dataset is serviced from this device.
Figure~\ref{fig:motivation-lsm-components} depicts the performance of RocksDB under write- and read-intensive workloads for all storage setups.

\paragraph{Write-intensive workloads}
Placing all LSM components in the faster storage device (ext4-DAX and H$_5$ setups) increases RocksDB throughput up to 1.4$\times$ and 3.7$\times$ when compared to ext4 with NVMe and SATA SSDs, respectively.
Also, the custom storage backend improves throughput up to 20\% over ext4-DAX due to its PMDK driver, which stores and accesses data items (both I/O and metadata) directly from user-space.
However, maybe surprisingly, we observe that \emph{placing the entire dataset on \pmem is actually detrimental}, degrading throughput up to 35\% when compared to just placing a small subset of data items (H$_4$).
The reason behind this performance decrease is a direct result of the increased write concurrency placed over \pmem, which exceeds the maximum parallelism supported by the device (\cref{subsec:motivation-device-parallelism}).
Placing more components on \pmem results in more workers concurrently writing to it, which are originated from foreground writes to C$_{log}$, flushing C$_m$ to L$_0$, and multiple parallel compactions occurring over different levels of the LSM tree. 

\begin{takeaway}
    Tiering is the recommended strategy to improve write performance in LSM-based KVS.
    LSM components closer to C$_m$ must be placed on faster storage tiers.
    Not only because these handle the critical write data path, but also because placing more LSM levels degrades overall performance due to the poor write concurrency of emerging storage devices.
\end{takeaway}
\vspace*{2.5pt}

\paragraph{Read-intensive workloads}
Contrary to write-intensive workloads, RocksDB performance improves with the number of components placed on \pmem.
Specifically, ext4-DAX and H$_5$ surpass ext4 throughput by up to 2.3$\times$.
This is because while \pmem's writes do not scale well, reads can scale up to the number of physical cores (\cref{subsec:motivation-device-parallelism}).
As such, \emph{placing the entire dataset in the faster tier of the storage hierarchy is beneficial for read-intensive workloads}.

\begin{takeaway}
    Caching is the recommended strategy to improve read performance in LSM-based KVS.
    Leveraging the fact that emerging storage devices offer excellent read bandwidth and scalability, one should strive to maximize their hit ratio. 
\end{takeaway}

\begin{figure}[t]
    \centering
    \includegraphics[width=1.0\linewidth,keepaspectratio]{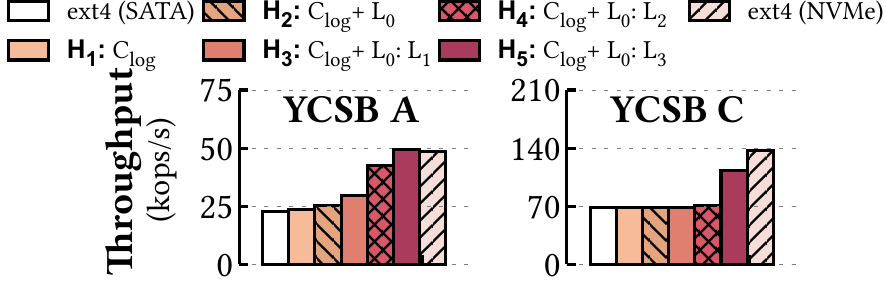}
    \caption{RocksDB performance when placing LSM components across an NVMe-SATA SSDs hierarchy under mixed and read-only workloads.}
    \label{fig:revision-study-nvme-sata}%
    \vspace*{-10pt}
\end{figure}

\revision{\paragraph{Performance under block-based storage hierarchies}
To understand whether the previous observations generalize to block-based storage hierarchies, we explored the impact of placing LSM components across \nvme and \sata devices.
We measured RocksDB's throughput under ext4 with SATA and {\nvme}s, as well as with the custom storage backend configured with different placement schemes.
We used 16 client threads and 16 background threads (exceeding NVMe's parallelism limit, as observed in \cref{subsec:motivation-device-parallelism}) under a 400~GiB dataset.
For the custom storage backend, the NVMe device capacity was limited to 100~GiB. 
To minimize caching effects, we disabled the OS page cache. 
Figure~\ref{fig:revision-study-nvme-sata} depicts the performance of all storage setups under write- and read-heavy workloads with a uniform key distribution. 
While concurrency effects are not as pronounced compared to an \pmem-NVMe storage hierarchy, we observe similar performance trends. 
For write-heavy workloads, placing the entire dataset on NVMe offers no significant advantage over placing only a small portion of the LSM ($H_5$).
For read workloads, RocksDB performance improves as more LSM components were placed on \nvme, aligning with the previous observations for hierarchies with \pmem devices.}

\begin{takeaway}
    \revision{Placement-aware optimizations that account for device-level parallelism, I/O bandwidth, and capacity remain relevant even in block-based storage hierarchies.}
\end{takeaway}


\subsection{Concurrent compactions}
\label{subsec:motivation-concurrent-compactions}

\begin{figure}[t]
    \centering
    \includegraphics[width=1\linewidth,keepaspectratio]{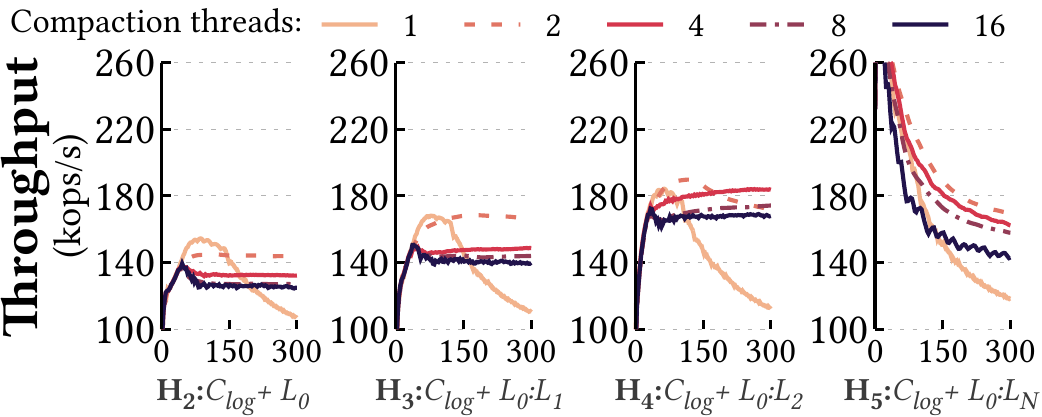}
    \caption{RocksDB performance over time (in seconds) under different placement schemes with increasing number of concurrent compactions. The caption of each plot depicts the components placed on \pmem.}
    \label{fig:motivation-compactions-concurrency}
\end{figure}

We now analyze the performance impact of increasing the number of writers and LSM components stored across the storage hierarchy. 
We fixed the number of client (8) and flushing (1) threads but varied the number of compaction threads (1 to 16).
Figure~\ref{fig:motivation-compactions-concurrency} shows the throughput over time across all configurations under YCSB A.

When a small amount of data is persisted in \pmem (namely, H$_2$ and H$_3$), RocksDB's performance is suboptimal since a large amount of requests are serviced by the \nvme.
Increasing the number of threads does not improve performance since the problem lies in the placement scheme.
Similarly to \cref{subsec:motivation-lsm-placement}, RocksDB achieves the best throughput performance under the H$_4$ configuration, when using 4 compaction threads.
With a low number of threads (\emph{i.e.,} up to 2), RocksDB experiences write stalls, which occur when flushes and low-level compactions are slow or on hold~\cite{SILK:2019:Balmau,ADOC:2023:Yu}.
On the other hand, too many compaction threads increases the number of writers contending the \pmem.
These effect becomes further accentuated when all components reside on the faster tier (H$_5$).

\begin{takeaway}
    The KVS performance is directly impacted by the placement of components across the hierarchy and the number of writers.
    This means that there is no single placement scheme that fits all workloads and system configurations.
\end{takeaway}


\subsection{Popularity of LSM levels}
\label{subsec:motivation-level-popularity}

We now investigate how different data distributions impact the storage hierarchy.
We analyze the accesses over time across the LSM (levels and block cache) under different distributions, including \emph{zipfian 0.99} (high skew), \emph{zipfian 0.80} (medium skew), and \emph{uniform}.

\begin{figure}[t]
    \centering
    \includegraphics[width=1\linewidth,keepaspectratio]{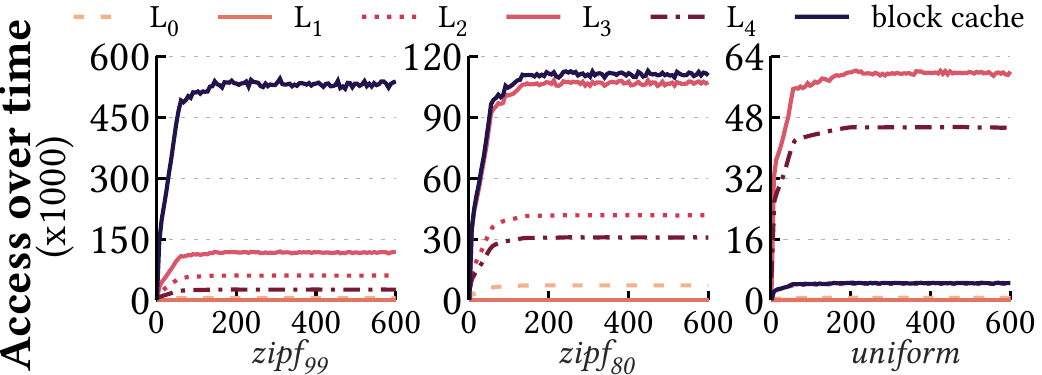}
    \caption{Read operations done over time (in seconds) across LSM levels under different data distributions.} 
    \label{fig:motivation-access-patterns}
    \vspace*{-15pt}
\end{figure}

\paragraph{Read workflows}
Figure~\ref{fig:motivation-access-patterns} depicts the number of foreground reads observed at each LSM level for a read-only workload. 
We observe that L$_3$ is the most accessed level.
While L$_0$ to L$_2$ have a combined size of approximately 3~GiB, L$_3$ holds 25~GiB of the dataset.
On the other hand, even though L$_4$ can accommodate 10$\times$ more data than L$_3$, it holds the older and colder portion of the dataset~\cite{LSM:2020:Luo,Mutant:2018:Yoon}.
We also observe that the block cache places an important role in highly skewed workloads, being able to service a large amount of reads, but its impact fades as the distribution becomes less skewed.
This means that for low skewed read-intensive workloads, levels higher than L$_2$ should be stored in the faster tier as well; otherwise, a significant portion of requests will be made over the \nvme.

\paragraph{Write workflows}
Due to the design of LSM KVS, the write data path follows the same workflow regardless of the data distribution. 
As such, components that are on the critical data path (\emph{i.e.,} C$_{log}$, L$_0$, L$_1$) should be placed on the faster tier, while the placement of other levels must be chosen to prevent write contention.

\begin{takeaway}
    When considering different data distributions, there is no winning placement strategy that simultaneously fits the requirements of read and write workflows.
\end{takeaway}

\section{\SYS storage middleware}
\label{sec:design}

We propose \SYS, a concurrency- and workload-aware storage middleware that accelerates the performance of existing LSM KVS running on a hierarchy of storage devices.
Through its design, \SYS automatically capitalizes on the strengths of the devices and compensates for their weaknesses, all while requiring minimal changes to the KVS.
Following the takeaways discussed in \cref{sec:study}, \SYS's design is built following five core principles.

\paragraph{Parallelism, bandwidth, and capacity-aware LSM placement}
\SYS realizes the full potential of the storage hierarchy by placing LSM components according to the \emph{parallelism}, \emph{bandwidth}, and \emph{capacity} of each storage medium.

\paragraph{Maximize hit ratio on faster devices}
\SYS exploits the high read bandwidth, scalability, and low latency of faster storage devices (\emph{e.g.,} \pmem), maximizing the number of requests served from them and minimizing read latency across the hierarchy.

\paragraph{Automatic partitioning and tuning}
To adapt to the performance characteristics of the different storage mediums, \SYS automatically manages the LSM partitioning across the hierarchy and the concurrency of background data movements between devices.

\paragraph{Flexibility and extensibility}
\SYS supports different combinations of storage devices and enables system designers to implement custom storage drivers and placement policies to comply with their workloads and system requirements.

\paragraph{Low intrusiveness}
\SYS requires minimal changes to LSM KVS, minimizing the work needed to maintain and port it to new systems.


\subsection{\SYS design overview}
\label{subsec:design-overview}

\begin{figure}[t]
    \centering
    \includegraphics[width=1\linewidth,keepaspectratio]{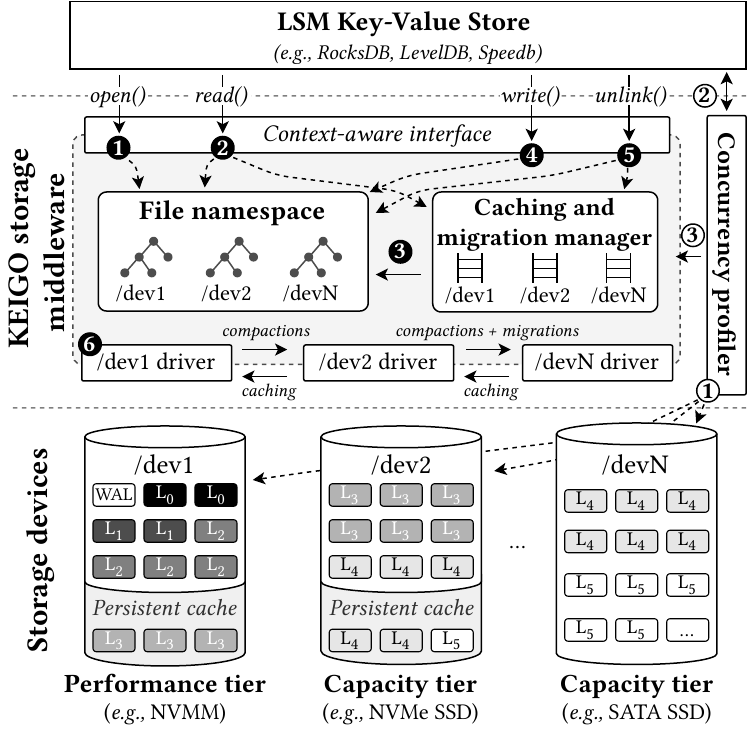}
    \caption{\textbf{\SYS architecture.} It follows a multi-tier storage hierarchy, co-designed with the properties of LSM KVS and the inherent characteristics of different storage devices.}
    \label{fig:sys-overview}%
    \vspace*{-10pt}
\end{figure}

Figure~\ref{fig:sys-overview} provides a high-level view of \SYS.
The system is a user-level middleware that sits between the LSM KVS (\emph{e.g.,} RocksDB, LevelDB) and a hierarchy of heterogeneous storage devices.
Building on the observation that placing as many KVS files as possible on the fastest storage tier can simultaneously improve read performance but degrade write performance, \SYS makes placement decisions that account for \emph{device-level parallelism}. 
To optimize writes, \SYS restricts placement on the faster device to only the LSM components in the critical data path, ensuring that performance-sensitive operations benefit from high throughput and low latency while keeping the number of active concurrent writers within the device's supported parallelism.
To optimize reads, \SYS asynchronously caches hot SST files on faster tiers of the hierarchy, lowering read latency without interfering with write performance. 

\SYS is organized as follows. 
At its core, a \emph{file namespace} component provides a logical-to-physical mapping of KVS files (\emph{e.g.,} SST, C$_{log}$) to their corresponding location, transparently abstracting a hierarchy of storage devices into a single logical one.
The hotness of each file is continuously tracked through a \emph{caching and migration manager}, which decides when to move data across devices either to improve the KVS performance (\emph{caching}) or for space management (\emph{migrations}).
To enable combining different types of storage devices, \SYS exposes a \emph{driver interface} that allows implementing custom I/O logic for each device, while supporting specialized interfaces and I/O protocols (\emph{e.g.,} PMDK~\cite{pmdk}, io\_uring).
In this paper, we assume a storage hierarchy made of byte-addressable (\pmem) and block-addressable storage devices (NVMe and SATA SSDs), acting as \emph{performance} and \emph{capacity devices}, respectively. 
DRAM plays a minimal role in \SYS's design, as most LSM KVS already perform internal memory management (\emph{e.g.,} block cache, memtable) to optimize operations in the critical data path.
To automate the partitioning of files across the hierarchy, \SYS integrates an \emph{offline concurrency profiler} that generates placement schemes based on the performance and parallelism of the storage devices in the hierarchy, as well as the KVS workload. 
In sum, \SYS is responsible for managing how files are efficiently stored across a hierarchy of devices, while other responsibilities, such as thread management and compaction scheduling, remain under the control of the KVS.

\noindent
\SYS is driven by three main techniques:

\emphparagraph{\textbf{1) Concurrency-aware data placement (\cref{subsec:optimizing-writes})}}
On-disk LSM components are split across the storage hierarchy based on two placement policies, driven simultaneously by the properties of the LSM and by the parallelism, I/O bandwidth, and capacity of each storage device.
With \emph{performance-aware placement}, LSM components whose operations are in the critical data path (C$_{log}$ and levels closer to C$_m$) and whose number of concurrent writers does not exceed the device's supported parallelism are placed on the faster storage device (\texttt{/dev1}).
With \emph{capacity-aware placement}, the bottom LSM-tree levels, which accommodate the colder majority of data and is where non-critical work is conducted, are placed on the remainder devices (\texttt{/dev2}, ..., \texttt{/devN}).

\emphparagraph{\textbf{2) Persistent, read-only caching (\cref{subsec:optimizing-reads})}}
Read-heavy workloads can experience low performance if a substantial amount of requests are handled by slower devices in the hierarchy. 
\SYS integrates a per-device persistent read-only cache that tracks and copies hot SST files from slower to faster storage devices, maximi\-zing the hit ratio on the upper levels of the hierarchy and improving the KVS performance under low skewed, read-intensive workloads.\footnote{The current \SYS prototype caches full SST files, but it is possible to further refine the policies for block-level caching.}

\emphparagraph{\textbf{3) Context-based I/O differentiation (\cref{subsec:context-propagation})}}
To minimize the intrusiveness of porting existing KVS, \SYS exposes a POSIX-like interface where system calls are passed with an additional \emph{context} field. 
The \emph{context} defines the internal KVS operation that originated a given POSIX request (\emph{e.g.,} flush, C$_{log}$, L$_N$ compaction), and is used to determine the device that will handle it. 
This is achieved by using \emph{context propagation}~\cite{ContextPropagation:2018:Mace,PAIO:2022:Macedo}, enabling KVS-level information to be propagated to \SYS with minor code changes.

\paragraph{Operation flow}
Figure~\ref{fig:sys-overview} illustrates how \SYS handles KVS operations. 
KVS files that contain clients' data are created when writing to C$_{log}$, flushing C$_m$ to disk, and during compactions. 
These operations are submitted to \SYS via extended \texttt{open()} calls that complement the standard POSIX interface with the \emph{context} that triggered such operation at the KVS level, whether it is a \texttt{log} write, \texttt{flush}, or \texttt{compaction} (and respective levels involved) (\ding{202}) (\cref{subsec:context-propagation}).
Based on this context, \SYS creates the file in the corresponding storage device according to the placement scheme (\ding{207}), generated offline by the \emph{concurrency profiler} (\cref{subsec:optimizing-writes}). 
To transparently support POSIX I/O to the KVS across a storage hierarchy, \SYS's file namespace maintains a logical-to-physical mapping of each file, linking the SST file with its physical location, original file identifier (\emph{i.e.,} file descriptor, memory address), and logical file descriptor returned to the KVS, which is used in subsequent I/O operations to that file such as \texttt{read()}, \texttt{write()}, and \texttt{close()}.
%
Writes to the critical data path (\ding{205}), coming from C$_{log}$, flush, and low-level compactions, are handled by the faster storage device (\texttt{/dev1}), while foreground reads are served from the tier where the file currently resides, whether cached or placed by the policy (\ding{203}).
Files may be moved between devices during compactions involving levels placed at different tiers (\cref{subsec:optimizing-writes}); when \emph{caching} hot files to faster devices in the hierarchy (\cref{subsec:optimizing-reads}); and during \emph{migrations} triggered for space management (\ding{204}) (\cref{subsec:optimizing-writes}).
When a compaction completes, unlinked files are removed from the caching/migration manager and the file namespace (\ding{206}).


\subsection{Optimizing writes} 
\label{subsec:optimizing-writes}

\SYS integrates the takeaways from \cref{sec:study}, by providing two placement policies to optimize writes.

\paragraph{Performance-aware data placement}
The \emph{performance placement policy}, applied to the fastest device in the storage hierarchy (\texttt{/dev1}), manages the LSM components that directly impact the performance perceived by clients.
The policy follows two key rules: \emph{1)} identify the components that lie in the critical data path that must be placed on the faster tier to minimize latency, and \emph{2)} determine the additional components that can be placed without exceeding the device's supported write parallelism and capacity.

\emphparagraph{1) Performance-critical components are placed in the faster tier}
\SYS places C$_{log}$ and lower levels of the LSM (L$_0$ and L$_1$) on the faster tier.
Our reasoning is threefold.
First, writes to C$_{log}$ need to be fast since they incur significant overhead to the critical data path, especially when the OS page cache is bypassed (\emph{e.g.,} \texttt{O\_DIRECT}) or when crash-consistency guarantees are desirable. 
Second, LSM workloads have strong temporal locality, where the popularity of objects fades over time~\cite{Mutant:2018:Yoon,FacebookTraces:2020:Cao}.
This means that the most accessed keys are stored in the most recent SST files, which are placed at the lower LSM levels.
Third, background tasks that involve C$_m$ and the lower levels of the tree are prone to write stalls, especially under write-heavy workloads. 
Stalls occur when flushes cannot proceed, either because flushes and low-level compactions are slow or on hold, resulting in degraded performance~\cite{SILK:2019:Balmau}. 
While placing these components on the faster tier does not avoid write stalls, their duration and performance degradation are hampered~\cite{ADOC:2023:Yu}.

\emphparagraph{2) Ensure the maximum device-level parallelism is not exceeded}
I/O bandwidth cannot serve as the sole determining factor for placing files in the faster tier, as the KVS performance under write-intensive workloads is significantly impacted by the number of active writers in the system, especially on \pmem devices (\cref{sec:study}). 
This is particularly detrimental for higher levels of the LSM, which perform larger compactions and contend the device for longer periods~\cite{SILK:2019:Balmau}.
As such, \SYS ensures that levels placed in the faster device are not solely determined by its bandwidth and space but also by the degree of parallelism among writer threads (\emph{i.e.,} writes to C$_{log}$, flushes, parallel compactions). 
This means that under write-heavy workloads, by selectively storing a small subset of KVS files (\emph{e.g.,} C$_{log}$ + L$_0$ to L$_2$, with a combined size of $\approx$3~GiB (\cref{subsec:motivation-level-popularity})) on the faster device, \SYS reduces write contention and improves performance. 

\emptyparagraph

\SYS implements these rules through an \emph{offline concurrency profiler} that generates a placement scheme with the LSM levels that must be placed in the faster storage tier.
The profiling process is made in three phases, as depicted in Figure~\ref{fig:sys-overview}.
First, it profiles the performance of each storage device, using the \texttt{fio} benchmarking tool~\cite{fio}, under read and write workloads with increasing number of threads until performance degradation is observed. 
This allows determining the maximum parallelism supported by each device (\cref{subsec:motivation-device-parallelism}). 
Second, it profiles the average number of concurrent writes that occur at different levels of the LSM during write-heavy workloads, using the YCSB benchmark~\cite{YCSB:2010:Cooper} (\ding{193}). 
We used YCSB as it generates workloads representative of those evaluated in \cref{sec:evaluation}. 
Nevertheless, system designers can use alternative configurations and workloads that more accurately reflect their production environments.
Based on these results, the profiler generates a scheme with the LSM levels whose cumulative concurrency demand does not exceed the supported write parallelism of the faster tier (\ding{194}).
Similarly to prior work~\cite{PrismDB:2023:Raina,Prism:2023:Song,MatrixKV:2020:Yao,SpanDB:2021:Chen}, \SYS assumes the internal state of the LSM remains stable over time (\emph{e.g.,} size of LSM levels, number of active threads).
While our experiments validate the effectiveness of this approach, there may be scenarios where the data placement should change over time (\emph{e.g.,} shifting access patterns, varying number of writers).
We leave the use of \emph{online profiling} for future work.

\paragraph{Capacity-aware data placement}
In the \emph{capacity placement policy}, applied to the remainder devices (\texttt{/dev2}, ..., \texttt{/devN}), LSM levels are placed in each tier according to their available storage space.
Our reasoning is that as new objects are inserted or updated in the KVS, older values are pushed down the stack and stored at the bottom LSM levels, thus accommodating the \emph{main bulk} and \emph{colder} portion of the dataset. 
To ensure sufficient capacity for incoming files, SST files are migrated across devices following an LRU eviction scheme.
Moreover, due to the size of the bottom LSM levels (\emph{i.e.,} hundreds of GBs to TBs), levels can be stored across multiple tiers.

\paragraph{Data movements across the hierarchy}
In \SYS, data operations between devices arise from two sources: 
compactions between LSM levels placed in different devices (managed by the KVS) and explicit operations performed by \SYS for performance and space management, including caching (\cref{subsec:optimizing-reads}) and migrations.
For compactions, \SYS performs a standard logical-to-physical mapping mechanism, translating the POSIX calls submitted by the KVS into their corresponding device-specific accesses -- namely, reads SST files from the targeted levels, writes the newly generated ones in the corresponding location, and removes obsolete SST files. 

For migrations, SST files are managed with an LRU eviction policy based on file access frequency. 
Candidate SST files for eviction are placed in a dedicated queue and moved by a thread pool once an upper-bound threshold is exceeded. 
This threshold ensures migrations are only triggered when device utilization surpasses a certain limit, avoiding premature eviction of SST files and minimizing inter-device traffic.
To avoid migrating SST files from upper LSM levels, particularly during atypical bursts to older key-value pairs, the eviction policy weighs the level at which SSTs respect to, prioritizing the migration of files from deeper (\emph{i.e.,} colder) LSM levels.
Further, to prevent the write concurrency problem observed in \cref{sec:study}, the \emph{caching and migration manager} component tracks the number of active writers in each device (\emph{i.e.,} compactions, caching, ongoing migrations) through atomic counters and dynamically adjusts migration parallelism to stay within the device’s supported write parallelism.
Finally, to reserve space for incoming SST files, \SYS forces migration when free space falls below a lower-bound threshold.\footnote{In our experiments, the upper- and lower-bound thresholds were set to 5\% and 2\% of the device capacity, respectively, but these values are user-configurable parameters.} 
While this could, in theory, temporarily exceed the device's supported parallelism, we have not observed this in practice; nevertheless, since capacity devices store the non-critical portion of the dataset, we do not expect any noticeable impact.

\subsection{Optimizing reads} 
\label{subsec:optimizing-reads}

Under read-heavy workloads exhibiting medium skew or uniform distributions (\cref{subsec:motivation-level-popularity}), most requests are made over the slower devices in the hierarchy (\texttt{/dev2}, ..., \texttt{/devN}).
While placing more levels on \texttt{/dev1} improves the performance of read-dominated workloads, it would severely impact write-heavy ones (\cref{subsec:motivation-lsm-placement}).
\SYS overcomes this challenge by reserving space at each storage device to persistently cache frequently accessed SST files from devices at the lower levels of the hierarchy.
For example, hot SST files from device \texttt{/devY} (\emph{e.g.,} \texttt{/dev2}) are temporarily cached on \texttt{/devY-1} (\emph{e.g.,} \texttt{/dev1}) as \emph{read-only copies}, while the \emph{original} files remain persisted in \texttt{/devY}.
Cached files are created in read-only mode, as in traditional LSM KVS (\emph{e.g.,} RocksDB) files become immutable after being fully written~\cite{SpaceAmplificationRocksDB:2017:Dong}.
Files are removed from the cache when the \emph{original} file is deleted (\emph{e.g.,} compactions) or when space is needed for hotter SST files.
\SYS' caching process is applied to all devices in the hierarchy, except for the last one, and addresses the following questions: 
\emph{1)} \emph{which} files should be cached; \emph{2)} \emph{when} should files be cached; and \emph{3)} \emph{how} should the actual process of caching be made in the presence of workloads with different read-write proportions.

\paragraph{\emph{1) File temperature profiling}}
\SYS determines \emph{which} files should be copied to the persistent cache by tracking SST access frequency.
Access counters are maintained at the file namespace and are updated by the KVS foreground threads (during reads that miss the block cache). 
To prevent caching stale data due to shifting access patterns or compactions, \SYS applies an aging factor to each SST file, decreasing the access counter in an exponential back-off manner each time a file is copied.

\paragraph{\emph{2) Hit ratio maximization}} 
\SYS determines \emph{when} to cache SST files by monitoring the hit ratio of foreground reads in the storage device that hosts the cache (\texttt{/devY-1}).
To achieve this, a dedicated background thread continuously computes the hit ratio of \texttt{/devY-1} and triggers a copy when the value is below a certain threshold.
When triggered, \SYS selects the most frequently accessed SST file from \texttt{/devY} and places it in an internal queue to be copied.
The monitoring granularity and hit ratio threshold are user-configurable. 

\paragraph{\emph{3) Concurrency-aware copying}}
\SYS uses a dedicated thread pool to cache files in parallel. 
However, this operation must be done carefully to avoid exacerbating the write concurrency problem observed in \cref{sec:study}.
As such, \SYS tracks the number of \emph{active writers} on the storage device that hosts the cache (\emph{e.g.,} \texttt{/dev1}) --- namely,  C$_{log}$, flush, compactions, and caching --- and dynamically adjusts the number of caching threads to ensure the total writers do not exceed the device's maximum parallelism level. 
The number of writers decreases when the corresponding background tasks finish.

Furthermore, in write-only workloads, the number of caching threads may temporarily drop to zero due to frequent flushes and compactions. 
This behavior is not detrimental, as caching is unnecessary in the absence of read operations.


\subsection{Context-based I/O differentiation} 
\label{subsec:context-propagation}

General-purpose storage systems are agnostic of the LSM's internal I/O logic, treating all requests in the same manner regardless of their priority and performance cost.
On the other hand, building specialized LSM KVS from the ground up to support heterogeneous storage devices requires significant implementation efforts, which are continuously repeated upon the release of new storage hardware.
\SYS strikes a balance between the two approaches by propagating the origin of KVS operations (\emph{e.g.,} C$_{log}$, flush, L$_0$ to L$_1$ compaction) to the storage, enabling the same level of control and performance as LSM-specific optimizations while imposing minimal code changes.
It combines ideas from \emph{context propagation}, a commonly used technique that enables a system to forward additional request information along its execution path~\cite{PAIO:2022:Macedo,ContextPropagation:2018:Mace,SplitIOScheduling:2015:Yang}, and applies them to determine which device should handle each request. 
The process for differentiating requests in \SYS is twofold.

\paragraph{Instrumentation}
First, \SYS needs minimal instrumentation on the data path of the LSM foreground and background work, including C$_{log}$, flushes, and compactions at different levels.
Whenever these operations are triggered, a tag (or \emph{context}) associated with the corresponding operation type (\emph{e.g.,} \texttt{log}, \texttt{flush}, \texttt{comp\_l0\_l1}) is registered at a variable residing in the local address space of each thread through the OS's thread-local storage mechanism~\cite{Memory:2007:Drepper}. 

\paragraph{Execution}
During execution, when a new file is created, an extended \texttt{open()} system call is sent to \SYS with the original arguments of POSIX \texttt{open()} and the \emph{context} that originated the request. 
This \emph{context} determines the device that will persist the file, according to the data placement scheme. 
Subsequent file requests (\emph{e.g.,} \texttt{read()}, \texttt{write()}, \texttt{close()}) do not need to pass the operation context, as they can access the file through \SYS's file namespace.


\subsection{Implementation} 
\label{subsec:implementation}

We implemented a prototype of \SYS in 4K LoC in C++. 
\SYS is provided as a user-level library, extending 25 POSIX calls, such as \texttt{open()}, \texttt{read()}, \texttt{pwrite()}, etc; we found that supporting this set of calls is sufficient to enable \SYS over LSM-based KVS (\cref{subsec:evaluation-portability}).

\paragraph{Storage drivers}
We implemented two storage drivers: a \pmem driver for managing byte-addressable and a POSIX driver for block-based storage devices. 
The former is implemented using Intel PMDK.
POSIX operations are converted into their corresponding memory-mapped version: \revision{\texttt{open()} and \texttt{close()} system calls are handled using \texttt{pmem\_map\_file()} and \texttt{pmem\_unmap()} routines,} reads are serviced via \texttt{memcpy()}, and writes with \texttt{pmem\_memcpy()} using \texttt{ntstore} instructions to avoid polluting the processor cache.
\SYS provides similar consistency guarantees as POSIX, making data durable upon explicit synchronization through \texttt{pmem\_flush()} and \texttt{pmem\_drain()} instructions.
For the POSIX driver, requests are submitted following standard POSIX semantics.


\paragraph{In-memory structures}
The \emph{file namespace} is implemented in a concurrent hashmap that maps \emph{logical} file descriptors to the metadata objects that maintain information about the actual file (\emph{e.g.,} original file descriptor, pathname).
Entries are atomically updated at file creation/removal (\emph{i.e.,} flush, compactions, C$_{log}$) and when moving files across devices (\emph{i.e.,} trivial moves, caching, migrations).

\revision{\paragraph{Context propagation}
To propagate the internal KVS operation that originated a given POSIX request to \SYS, we use the OS's thread-local storage mechanism~\cite{Memory:2007:Drepper} by storing this information in a variable residing in the local memory space of each thread, eliminating contention or potential race conditions.}

\paragraph{Persistent caches}
Each \emph{persistent cache} uses a dedicated thread pool for background copying of SST files and separate threads for monitoring the hit ratio of each storage device. 
The hit ratio threshold is defined by a fixed, configurable parameter.
On file copying, \SYS prefetches targeted SST files to the OS page cache using \texttt{fadvise()} and \texttt{readahead()}. 
After its completion, it hints the OS to evict the pages that hold prefetched data.


\paragraph{Porting LSM-based KVS}
We integrated \SYS with three popular LSM KVS, namely RocksDB~\cite{rocksdb-git} (57 LoC), Speedb~\cite{speedb-git} (81 LoC), and LevelDB~\cite{leveldb-git} (40 LoC), which required updating the thread-local variables for context propagation and replacing POSIX system calls with those supported by \SYS.

\revision{\paragraph{Trivial moves}
At the LSM level, \emph{trivial moves} are logical operations that move SST files between levels without explicitly triggering a compaction.
However, in a storage hierarchy, files need to be physically moved if the target level resides on a different device. 
We instrumented RocksDB's data path responsible for trivial moves; whenever a trivial move is triggered, the corresponding file is placed in a queue where a dedicated thread pool performs the migration.}


\begin{figure*}[t]
    \centering
    \begin{subfigure}[t]{1\linewidth}
        \includegraphics[width=1\linewidth,keepaspectratio]{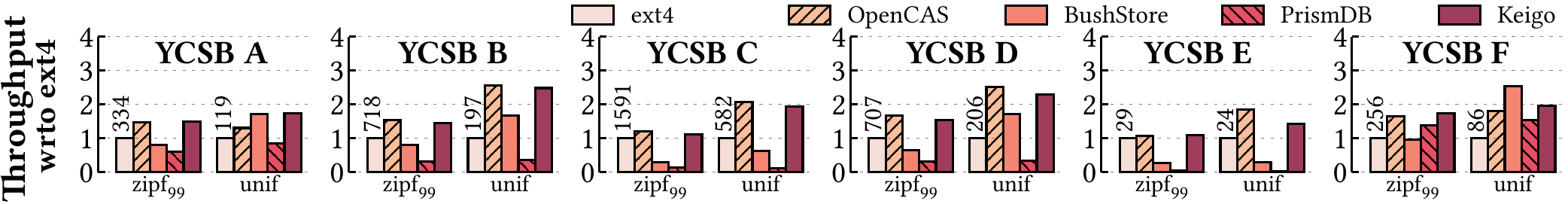}
        \caption{\small Relative throughput of all systems, with respect to \emph{ext4}, over a 50~GiB dataset under zipf99 and uniform data distributions.}
        \label{fig:ycsb-workloads-50gb}%
    \end{subfigure}
    \begin{subfigure}[t]{1\linewidth}
        \includegraphics[width=1\linewidth,keepaspectratio]{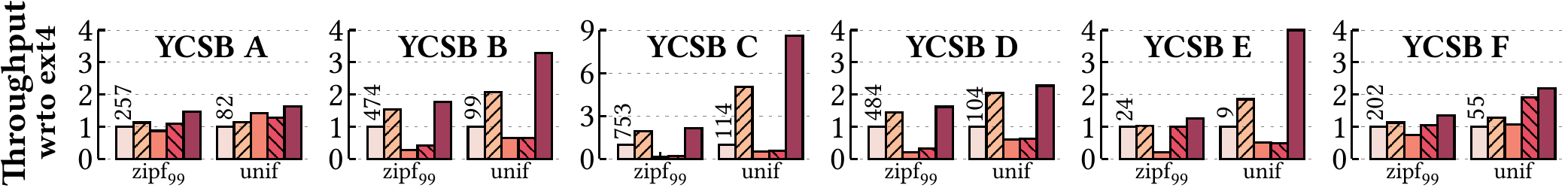}
        \subcaption{\small Relative throughput of all systems, with respect to \emph{ext4}, over a 200~GiB dataset under zipf99 and uniform data distributions.}
        \label{fig:ycsb-workloads-200gb}%
    \end{subfigure}
    \begin{subfigure}[t]{1\linewidth}
        \includegraphics[width=1\linewidth,keepaspectratio]{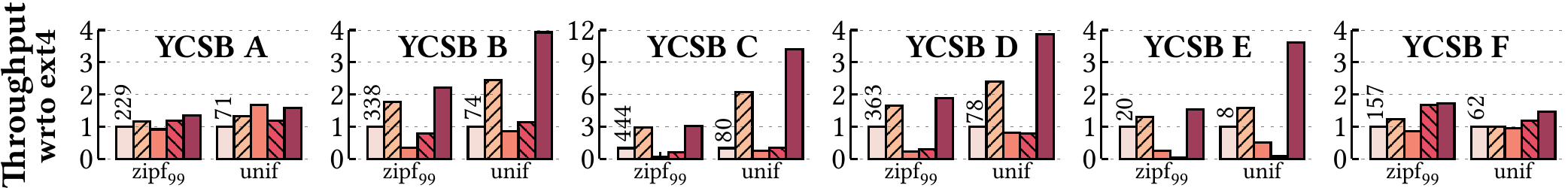}
        \caption{\small Relative throughput of all systems, with respect to \emph{ext4}, over a 400~GiB dataset under zipf99 and uniform data distributions.}
        \label{fig:ycsb-workloads-400gb}%
    \end{subfigure}
    \caption{Relative throughput of \emph{ext4}, \emph{OpenCAS}, \emph{BushStore}, \emph{PrismDB}, and \emph{\SYS} for YCSB workloads (A to F) under distinct data distributions and dataset sizes. 
    The absolute throughput value (in kops/s) of \emph{ext4} is shown above the bar.}
\end{figure*}

\subsection{Discussion} 
\label{subsec:discussion}

\paragraph{Impact of device-level parallelism}
Device-level parallelism is a fundamental characteristic that universally affects storage devices.
While its effects are especially pronounced in \pmem due to its inherent read-write asymmetry, our findings demonstrated that block-addressable devices like NVMe and SATA SSDs are also impacted (\cref{subsec:motivation-device-parallelism}), resulting in suboptimal performance under high write concurrency. 
In \cref{subsec:evaluation-sensitivity}, we demonstrate how \SYS improves LSM KVS performance even in block-addressable storage hierarchies.

\paragraph{Performance under multiple \pmem devices}
While our experiments focus on using a single \pmem, the findings and core principles of \SYS extend to setups with multiple devices.
Previous studies demonstrate that increasing the number of devices enhances overall write bandwidth but does not eliminate the concurrency limit inherent to each device~\cite{PmemDatabasesStudy:2023:Koutsoukos, PMMStudy:2020:Yang}, highlighting the importance of \SYS's concurrency-aware placement scheme.


\paragraph{Internal metadata consistency}
\SYS's current implementation is fault tolerant. 
Upon a crash, \SYS reconstructs its namespace by scanning the files stored in each device, which allows redirecting operations to the correct location.
Cached files may be present during recovery but do not affect \SYS's correctness, as they are removed during normal operation.
We defer the improvement of crash-recovery mechanisms for \SYS's metadata to future work.

\section{Evaluation}
\label{sec:evaluation}

Our evaluation sets out to answer the following questions:
\begin{itemize}[leftmargin=*]
    \item How does \SYS perform under varying dataset sizes and I/O guarantees (\cref{subsec:evaluation-datasets})?
    \item How does \SYS handle different levels of concurrency (\cref{subsec:evaluation-concurrency})?
    \item How does \SYS perform under production workloads (\cref{subsec:evaluation-production})?
    \item Can \SYS improve the performance of other LSM KVS (\cref{subsec:evaluation-portability})?
    \item What is the performance breakdown of the different techniques implemented in \SYS (\cref{subsec:evaluation-sensitivity})?
\end{itemize}

\paragraph{Hardware and OS configurations}
We used the same hardware and OS configurations as in \cref{sec:study}, with the exception of the \sata, which was replaced by two 480~GiB {\sata}s configured with RAID-0.
Unless stated otherwise, experiments were conducted using the \pmem and \nvme devices.

\paragraph{Baselines}
We evaluate and compare \SYS over two groups of systems: \emph{general-purpose storage systems}, namely ext4 and OpenCAS~\cite{OpenCAS}; and state-of-the-art \emph{LSMs designed for heterogeneous storage devices}, namely BushStore~\cite{BushStore:2024:Wang} and PrismDB~\cite{PrismDB:2023:Raina}.
\emph{\textbf{ext4}} corresponds to an ext4 file system mounted over \nvme.
\emph{\textbf{OpenCAS}} is an in-kernel Linux module that provides transparent and persistent caching to applications. 
We configured OpenCAS to use the \nvme as main storage (\emph{capacity device}) and \pmem as persistent cache following \emph{write-back} policy -- writes are submitted to the cache and acknowledged to the application before being written to the NVMe; periodically, these writes are flushed to the capacity device.
\emph{\textbf{BushStore}} is a LSM KVS that organizes L$_0$ and L$_1$ in a B+Tree structure and stores them in \pmem, while the remainder levels are stored on \nvme.
\emph{\textbf{PrismDB}} is a LSM KVS that stores hot data items (organized in slab files) and the C$_{log}$ on \pmem, and cold data items (in SST files) on NVMe.
The system also uses a multi-tiered compaction scheme to minimize I/O stalls.
As for \emph{\textbf{\SYS}}, unless stated otherwise, \pmem handles C$_{log}$ and L$_0$ to L$_2$, while the remainder levels are placed on \nvme, as generated by the offline concurrency profiler. 
Leftover \pmem space ($\approx$100~GiB) is allocated to the persistent cache. 
The thread pools for managing caching and migrations are configured with 16 threads.

\paragraph{Experimental setup}
\revision{Unless stated otherwise, systems were configured as follows.}
The ext4, OpenCAS, and \SYS experiments were conducted over RocksDB configured with two C$_m$ of 128~MiB each, a 1~GiB block cache, and a thread pool of 4 threads for internal operations, including 1 flushing thread. 
For optimal performance, BushStore was set with 64 threads for background work, and PrismDB with 1 background thread per client.
All experiments (except \cref{subsec:evaluation-concurrency}) use 8 client threads.
The data loading phase is single-threaded using a uniform distribution with 16B keys and 1024B values.
\revision{Experiments were conducted with workloads from the YCSB benchmark under the \emph{zipf99}, \emph{zipf80}, and \emph{uniform} key distributions.}


\subsection{Varying dataset sizes}
\label{subsec:evaluation-datasets}

\begin{figure}[t]
    \centering
    \includegraphics[width=1.0\linewidth,keepaspectratio]{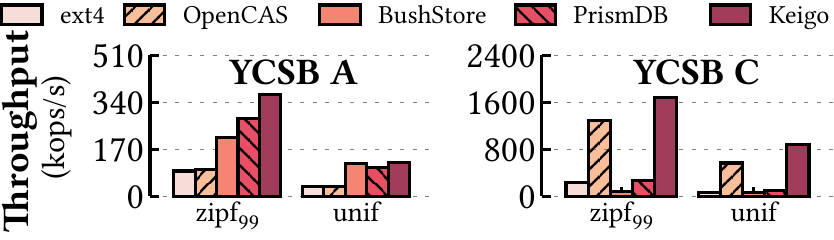}
    \caption{Performance under direct I/O for YCSB A and C.}
    \label{fig:ycsb-workloads-200gb-odirect}%
    \vspace*{-10pt}
\end{figure}

To understand how each system manages data across the storage hierarchy, we compare their performance under different dataset sizes, which can fit entirely (50~GiB) or only partially (200~GiB and 400~GiB) in the faster storage device (\emph{i.e.,} \pmem).
Experiments were conducted using YCSB workloads A to F.
Figures~\ref{fig:ycsb-workloads-50gb}, \ref{fig:ycsb-workloads-200gb}, and \ref{fig:ycsb-workloads-400gb} depict the relative throughput of all systems with respect to ext4 for 50~GiB, 200~GiB, and 400~GiB datasets, respectively.

\paragraph{Write-intensive workloads}
Under write-intensive workloads (A, F), \SYS outperforms all systems.
For datasets that do not fit entirely on \pmem, \SYS improves throughput up to 2.2$\times$, 1.7$\times$, 1.6$\times$, and 1.4$\times$ over ext4, OpenCAS, BushStore, and PrismDB.
The reason behind this performance difference is that in ext4, all requests are handled by the \nvme since it is the only device that can accommodate all datasets; 
OpenCAS caches requests in \pmem but the choice of which items should be cached is agnostic to the KVS, not considering their level or priority. 
The closest competitors are BushStore and PrismDB, which similarly to \SYS, handle the critical data path (\emph{i.e.,} C$_{log}$, L$_0$, L$_1$) on \pmem.
\SYS's performance difference becomes more pronounced as the dataset size increases. 
In \SYS, L$_2$ reads and writes are serviced by \pmem, while other solutions use the \nvme.

\paragraph{Read-intensive workloads}
Under read-intensive workloads (B, C, D), \SYS significantly outperforms all systems, especially under datasets that do not fit on \pmem.
\SYS show performance improvements of up to 10$\times$, 1.7$\times$, 18$\times$, and 15$\times$ over ext4, OpenCAS, BushStore, and PrismDB, a direct result of the persistent caching mechanism (\cref{subsec:optimizing-reads}).
In these experiments, ext4 and OpenCAS cache hot data items in the OS page cache and \pmem, being particularly noticeable under the zipf99 distribution.
For small datasets, OpenCAS demonstrates similar performance to \SYS, as it caches the entire dataset on \pmem.
On the other hand, BushStore and PrismDB experience poor performance due to their inability to cache hot data items either on \pmem or the OS page cache, leading to the majority of requests being serviced by \nvme.
Under scan-intensive workloads (E), \SYS improves performance up to 4$\times$ over general-purpose systems and 16$\times$ over the LSM KVS.

\begin{figure}[t]
    \centering
    \includegraphics[width=1.0\linewidth,keepaspectratio]{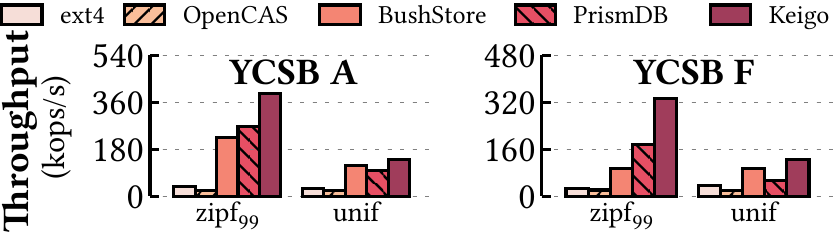}
    \caption{Performance under strict durability.}
    \label{fig:ycsb-workloads-200gb-fsync}%
    \vspace*{-10pt}
\end{figure}

\paragraph{Direct I/O}
In some production environments, applications use direct I/O to access their data, bypassing the OS page cache.
As such, we assess the performance of each system when the OS page cache is disabled (\texttt{O\_DIRECT}) under a 200~GiB dataset, depicted in Figure~\ref{fig:ycsb-workloads-200gb-odirect}.
\SYS, BushStore, and PrismDB demonstrate similar performance as with the OS page cache enabled, as they store the critical data path on \pmem with synchronous I/O to reduce CPU utilization by eliminating copies from the OS cache to the application buffer.
For YCSB A, \SYS overcomes ext4 and OpenCAS up to 4$\times$, as write operations are now directly submitted to the \nvme.
For YCSB C, \SYS overcomes ext4 by up to 12$\times$; OpenCAS exhibits similar performance as with the OS page cache enabled, since hot data items are eventually promoted to \pmem.

\revision{\paragraph{Strict durability}
Similarly to direct I/O, some applications require strict durability for crash consistency purposes~\cite{ByteHTAP:2022:Chen,Ceph:2019:Aghayev,LazyFS:2024:Ramos}.
We assess the performance of each system under strict durability settings by configuring the KVS to perform an \texttt{fsync()} call after each \texttt{write()} operation. 
Figure~\ref{fig:ycsb-workloads-200gb-fsync} depicts the results under a 200~GiB dataset using write-intensive workloads (YCSB A and F).
\SYS, BushStore, and PrismDB exhibit minimal performance degradation when compared to experiments without \texttt{fsync()} calls, as they now perform \texttt{pmem\_flush()} and \texttt{pmem\_drain()} instructions to flush data from CPU caches and wait for the memory controller to commit all writes to \pmem.
Moreover, under both workloads, \SYS overcomes ext4 and OpenCAS up to 17$\times$, as all operations are directly submitted to the \nvme bypassing all OS caches, including file system, block device, and device-level~\cite{LazyFS:2024:Ramos}.
}


\subsection{Varying concurrency levels}
\label{subsec:evaluation-concurrency}

\begin{figure}[t]
    \centering
    \includegraphics[width=1.0\linewidth,keepaspectratio]{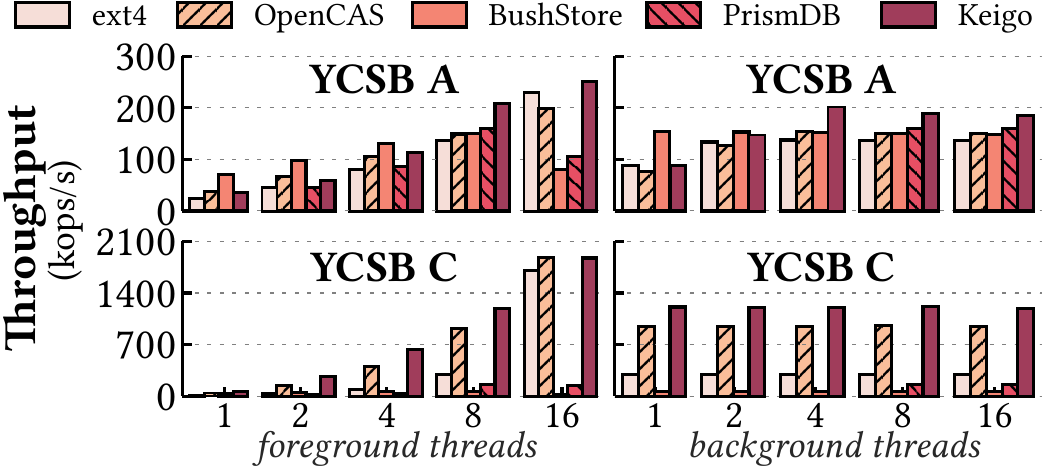}
    \caption{Performance of all systems for YCSB A and C with increasing number of foreground and background threads.}
    \label{fig:ycsb-concurrency}%
\end{figure}

Figure~\ref{fig:ycsb-concurrency} depicts the performance of all systems under varying concurrency levels.
Experiments were conducted for write-intensive (A) and read-intensive (C) workloads using a 200~GiB dataset under a zipf80 distribution with an increasing number of foreground and background threads, 
ranging from 1 to 16.
For the background experiments, all systems were configured with 8 client threads. 
We were unable to configure PrismDB for 1 to 4 background threads due to its requirement of having at least 1 thread per client.

\begin{figure}[t]
    \centering
    \includegraphics[width=1\linewidth,keepaspectratio]{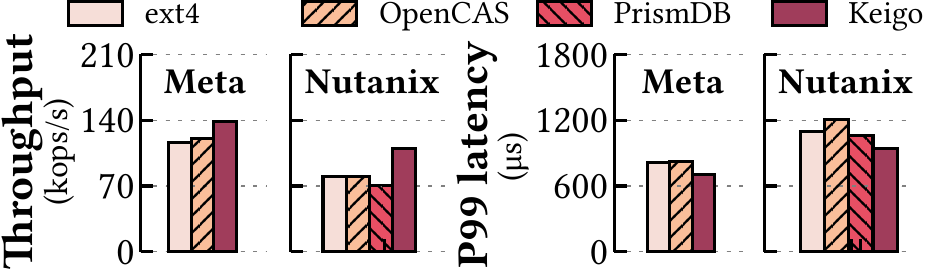}
    \caption{Performance of \emph{ext4}, \emph{OpenCAS}, \emph{PrismDB}, and \emph{\SYS} under Meta and Nutanix production workloads.}
    \label{fig:production-workloads}%
\end{figure}

\paragraph{Foreground concurrency}
Under the YCSB A workload, ext4, OpenCAS, and \SYS scale with the number of foreground workers. 
For a small number of workers (1 to 4), BushStore exhibits the best performance due to its B+Tree design under L$_0$ and L$_1$ LSM levels.
Interestingly, contrary to the other systems, both BushStore and PrismDB performance peaks at 8 client threads, and experience a performance decrease of 47\% and 34\% under 16 threads. 
The reasons behind this are that: for BushStore, since \pmem only stores the lower levels of the tree, the majority of requests are serviced by the \nvme device; for PrismDB, since it requires at least 1 background thread per client, the number of active writers in the system far exceeds the \pmem's limit (\emph{i.e.,} 4), causing the write concurrency problem observed in \cref{sec:study}.
For the read-intensive workload, we draw similar conclusions as in \cref{subsec:evaluation-datasets}.

\paragraph{Background concurrency}
\SYS was configured with different placement schemes generated by the concurrency profiler. 
Specifically, with up to 2 background threads, \SYS stored in \pmem all files up to L$_3$, while the other levels were placed on NVMe due to space constraints; for 8 and 16 threads, \SYS stored in \pmem all files up to L$_1$, as the device's supported write parallelism would be exceeded beyond that level. 
Under YCSB A, for a small number of threads, since ext4, OpenCAS, and \SYS use native RocksDB, all systems experience write stalls, as discussed in \cref{subsec:motivation-concurrent-compactions}.
Beyond 4 threads, \SYS achieves the best performance across all systems.
As for YCSB C, since it is a read-only workload (absent of compactions), all systems show stable performance with increasing number of workers.
Interestingly, while \SYS is configured with different placement schemes, its performance remains stable due to the persistent caching mechanism, achieving a performance increase of 4$\times$, 1.25$\times$, 18$\times$, and 7.5$\times$ over ext4, OpenCAS, BushStore, and PrismDB.


\subsection{Production workloads}
\label{subsec:evaluation-production}

We now evaluate how each system performs under production workloads from Meta~\cite{FacebookTraces:2020:Cao} and Nutanix~\cite{Kvell:2019:Lepers}.
Figure~\ref{fig:production-workloads} captures the throughput and tail latency of all systems.

\paragraph{Meta workloads}
We use the \emph{prefix\_dist} production workload from Meta~\cite{FacebookTraces:2020:Cao}, a read-dominated workload that combines read, write, and scan operations at a ratio of 83:14:3, uses varying key-value pair sizes, and exhibits realistic key hotness patterns.
Experiments ran over a 50M key-value pairs dataset using 8 client threads.
Due to its placement scheme and persistent cache, \SYS outperforms all baselines, increasing throughput by 1.2$\times$ over ext4 and 1.15$\times$ over OpenCAS.
We were unable to run BushStore and PrismDB under this workload, as they crash due to the varying key sizes.

\paragraph{Nutanix workloads}
The Nutanix workload is a write-intensive workload that combines read, write, and scan operations at a ratio of 40:58:2. 
The items requested during the execution also vary in size, ranging from 100B to 4~KiB, with a median size of 400B.
Experiments were conducted over a 500M key-value pairs dataset.
\SYS outperforms all baselines, achieving a throughput increase of up to 1.6$\times$ over other systems.
Similarly to the Meta workloads, we were unable to successfully run BushStore.


\subsection{Portability}
\label{subsec:evaluation-portability}

\begin{figure}[t]
    \centering
    \begin{subfigure}[t]{1\linewidth}
        \includegraphics[width=1.0\linewidth,keepaspectratio]{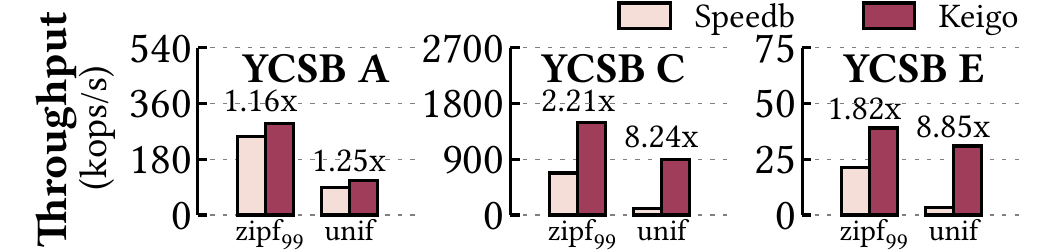}
        \vspace*{0.25pt}
    \end{subfigure}
    \begin{subfigure}[t]{1\linewidth}
        \includegraphics[width=1.0\linewidth,keepaspectratio]{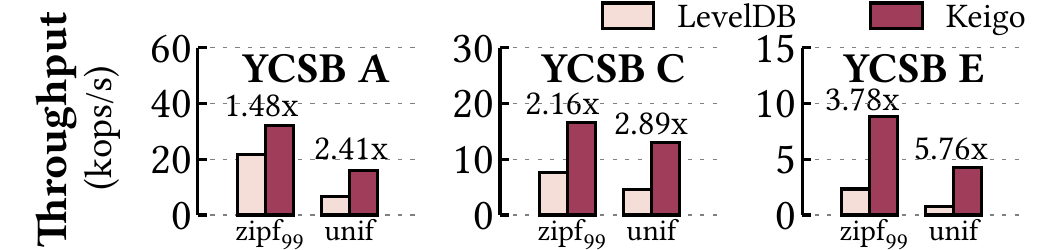}
    \end{subfigure}
    \caption{Throughput of Speedb and LevelDB with \SYS.}
    \label{fig:portability}%
    \vspace*{-10pt}
\end{figure}

We now demonstrate the portability of \SYS by integrating it with Speedb~\cite{speedb-git} (81 LoC) and LevelDB~\cite{leveldb-git} (40 LoC).
Figure~\ref{fig:portability} compares the performance of the base systems configured with the ext4 setup and their corresponding \SYS-enabled versions.
Experiments were conducted in a 200~GiB dataset with write- (A), read- (C), and scan-intensive (E) workloads under different data distributions.
Speedb experiments were conducted using 8 client threads, while LevelDB's were sequential due to the lack of concurrency support. 

\SYS improves the performance of both systems across all workloads due to its data placement and persistent caching optimizations.
Speedb performance is improved up to 8$\times$ for read- and scan-intensive workloads and up to 1.33$\times$ for write-intensive workloads.
As for LevelDB, \SYS shows a performance increase of up to 2.4$\times$ for YCSB A, 2.89$\times$ for YCSB C, and 5.76$\times$ for YCSB E.


\subsection{Sensitivity analysis}
\label{subsec:evaluation-sensitivity}

\begin{figure}[t]
    \centering
    \includegraphics[width=1.0\linewidth,keepaspectratio]{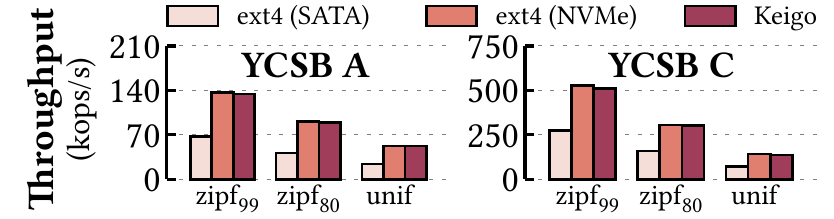}
    \caption{Performance of \SYS under a block-addressable storage hierarchy for YCSB A and C.}
    \label{fig:micro-nvme-sata}%
    \vspace*{-10pt}
\end{figure}

We now evaluate the performance of distinct features of \SYS. 

\paragraph{Block-addressable storage hierarchy}
We first study the performance of \SYS under a block-addressable storage hierarchy. 
The experiments were conducted for YCSB A and C workloads using a 200~GiB dataset with 16 client threads across three setups: ext4 mounted on the SATA SSD, ext4 mounted on the NVMe SSD, and \SYS using both NVMe and SATA SSDs.
For \SYS, the \nvme is limited to 100~GiB to simulate a scenario where the dataset exceeds the capacity of the faster tier.
As generated by the offline concurrency profiler, the \nvme handles $C_{log}$ and L$_0$ to L$_3$, while the remainder levels are placed on the \sata. 
Leftover NVMe space ($\approx$70~GiB) is allocated to the persistent cache.
To minimize caching effects, the OS page cache was disabled for all systems.

Figure~\ref{fig:micro-nvme-sata} depicts the performance of all setups over different data distributions.
Under an NVMe-\sata hierarchy, \SYS improves RocksDB performance by up to 2.2$\times$ for write-heavy workloads and 1.9$\times$ for read-heavy workloads compared to storing all LSM components on ext4 backed by the \sata.
When compared to a scenario where all LSM components reside on the \nvme device, \SYS achieves similar performance while placing only a small portion of the LSM ($\approx$30~GiB, 15\%) in the faster, and more expensive, tier. 
These results highlight \SYS's adaptability to different storage hierarchies and reinforce the broader applicability of its core principles (\cref{sec:design}) beyond \pmem-based storage hierarchies.

\begin{figure}[t]
    \centering
    \includegraphics[width=1.0\linewidth,keepaspectratio]{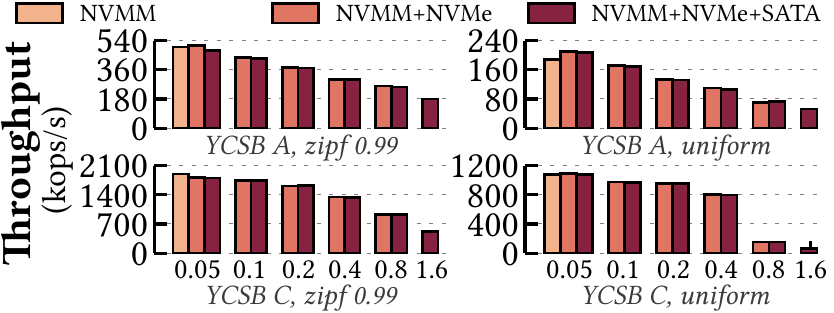}
    \caption{Performance with different tiers and dataset sizes.}
    \label{fig:micro-multi-tier}%
\end{figure}

\paragraph{Multiple storage tiers}
We now analyze the performance of \SYS under different storage hierarchies: a single tier composed of a \pmem; a tier using \pmem and \nvme devices; and the combination of \pmem, \nvme, and \sata.
Adding more devices to the hierarchy allows the system to handle larger datasets.
For these experiments, NVMe capacity was limited to 800~GiB, and the \pmem and NVMe's caches were configured to 80~GiB and 160~GiB, respectively.
Figure~\ref{fig:micro-multi-tier} depicts the performance of the different combinations for write- (A) and read-intensive (C) workloads over increasing dataset sizes, ranging from 50~GiB to 1.6~TiB.
All storage hierarchies achieve the best performance for smaller datasets, as the majority of requests are serviced by the faster tiers.
Interestingly, the \pmem+~\nvme and \pmem+~\nvme+~\sata hierarchies show similar performance under the same dataset, never exceeding a relative difference of 5\%.
This is due to \SYS's caching and migration mechanisms, which ensure sustained performance even when adding slower storage devices to the hierarchy.

\paragraph{Impact of the persistent cache}
We now analyze the performance impact imposed by \SYS's persistent cache.
We ran a read-intensive workload (C) for a 400~GiB dataset stored over a storage hierarchy composed of \pmem, \nvme, and \sata. 
We considered three setups: when caching is disabled, when \pmem cache is enabled, and when both caches (\pmem and \nvme) are enabled.
The storage capacity of each device was limited to 100~GiB, 200~GiB, and 400~GiB, respectively. 
The \pmem and \nvme's caches were configured to 70~GiB and 50~GiB, respectively. 

When caches are disabled, \SYS's performance ranges between 126~kops/s and 815~kops/s for uniform and zipf99 distributions.
When the \pmem cache is enabled, its throughput increases by 4.5$\times$ for uniform and 1.7$\times$ for zipf99.
Results also show that, under this dataset, the \nvme cache has a negligible impact on performance since most requests are serviced by the \pmem cache, either due to accesses over the hot SST files originally placed on \pmem or the hot files copied from the \nvme device.

\begin{figure}[t]
    \centering
    \includegraphics[width=1.0\linewidth,keepaspectratio]{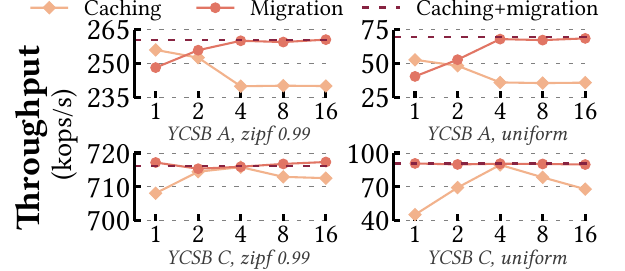}
    \caption{Performance of \SYS's automatic thread control for caching and migrations.} 
    \label{fig:micro-caching-eviction}%
\end{figure}

\paragraph{Caching and migration agressiveness}
As discussed in~\cref{subsec:optimizing-writes} and~\cref{subsec:optimizing-reads}, \SYS automatically controls the number of threads migrating and caching data across devices. 
Figure~\ref{fig:micro-caching-eviction} shows the performance of such mechanism for YCSB A and C workloads with an 800~GiB dataset. 
We compare three different setups: \emph{Caching} manually changes the number of caching threads from 1 to 16, while letting \SYS choose the number of migration threads; \emph{Migration} varies the number of migration threads and lets \SYS define the number of caching threads; and \emph{Caching+migration} lets \SYS automatically control the number of both thread types.

For write-intensive workloads, increasing the number of caching threads decreases the KVS throughput, as parallelizing the caching process (combined with ongoing writes from C$_{log}$, flush, and compactions) causes the number of active writers to exceed \pmem's concurrent writers limit.
For read-intensive workloads, since there are no active writers in the system, \SYS scales up to 4 caching threads.
As for migrations, since these are performed between the NVMe and SATA SSDs, the performance improves up to 16 threads.
When both caching and migrations operate with the automatic thread control mechanism, \SYS achieves the best performance across all experiments. 
This is because \SYS continuously monitors the number of active writers in each storage device and automatically adjusts the number of concurrent caching and migrations.

\section{Related work}
\label{sec:related-work}

This section describes prior work and places our work in context.

\paragraph{General-purpose hierarchical storage}
OpenCAS~\cite{OpenCAS} is a generic block-layer caching mechanism that enables using a fast device as a cache (for both reads and writes) of a slower device.
P2Cache~\cite{P2Cache:2023:Lin}, FirstResponder~\cite{FirstResponder:2021:Song}, and SPFS~\cite{SPFS:2023:Woo} are in-kernel caching mechanisms that enhance legacy file systems by using \pmem to absorb frequent writes. 
Orthus~\cite{Orthus:2021:Wu} introduces a non-hierarchical caching strategy that redirects requests based on device load.
Strata~\cite{Strata:2017:Kwon}, Ziggurat~\cite{Ziggurat:2019:Zheng}, and TPFS~\cite{TPFS:2023:Zheng} are file systems that tier data across DRAM, \pmem, and SSD according to the applications' access patterns and consistency requirements.
These systems, however, are agnostic of applications' internal I/O logic.
For LSM KVS, this means that requests with different priorities and storage performance costs are treated in the same manner across the storage hierarchy, impacting the KVS' end performance as observed in \cref{sec:study} and \cref{sec:evaluation}.
In contrast, \SYS is an LSM-aware middleware that places LSM components in the storage device that best suits their workload patterns.

\paragraph{KVS with hierarchical storage support}
Mutant~\cite{Mutant:2018:Yoon} places SST files based on their popularity over cloud-based storage while enforcing storage cost SLOs.
SpanDB~\cite{SpanDB:2021:Chen} uses high-performance SSDs via SPDK to store the C$_{log}$ and lower levels of the LSM tree.
PrismDB~\cite{PrismDB:2023:Raina} proposes a multi-tier compaction scheme that spawns across a hierarchy of heterogeneous storage devices.
MatrixKV~\cite{MatrixKV:2020:Yao} proposes a new matrix-like data structure that manages L$_0$ and is placed on \pmem.
BushStore~\cite{BushStore:2024:Wang} replaces traditional SSTs with B+Trees, placing L$_0$ and L$_1$ in byte-addressable storage, and the rest of the LSM in a block-based device.
WaLSM~\cite{WaLSM:2023:Chen} actively profiles data freshness and access frequency to accurately migrate cold data from \pmem to SSD.
Prism~\cite{Prism:2023:Song} is a KVS specifically designed for DRAM-\pmem-SSD storage hierarchies, that follows a key-value separation model, where keys are placed on \pmem and values are placed on SSD (first absorbed by \pmem to minimize write latency and later migrated to SSD). 
Prism introduces techniques that minimize thread synchronization over \emph{``wide''} storage hierarchies composed of multiple similar devices operating in parallel.  
Replacing LSMs used in production with these systems is not trivial, posing significant implementation efforts. 
\SYS is independent of specific LSM implementations or storage devices, enabling portability and performance improvements with minimal code changes.

Unlike previous works, \SYS is the first solution to fully leverage heterogeneous storage hierarchies by optimizing LSM I/O workflows for each device's \emph{parallelism}, \emph{bandwidth}, and \emph{capacity}. 
By doing so, \SYS significantly improves the performance of widely used LSM-based KVS systems under read and write workloads.

\revision{\paragraph{NVMM-based KVS}
Recent work has also explored the design of new KVS that exclusively use \pmem as main storage~\cite{HiKV:2017:Xia,NoveLSM:2018:Kannan,SLMDB:2019:Kaiyrakhmet,FlatStore:2020:Chen,ChameleonDB:2021:Zhang,NVLSM:2021:Zhang,Pacman:2022:Wang,ListDB:2022:Kim}, thus being orthogonal to \SYS.
These works are orthogonal to \SYS as they do not consider devices that reside at deeper levels of the hierarchy.}

\revision{\paragraph{Memory tiering}
Another line of research explores \pmem as a cheaper alternative to DRAM to design tiered memory systems~\cite{HeMem:2021:Raybuck,WritesHurt:2022:Fedorova,Tarmac:2022:Kim,MioDB:2023:Duan,JohnnyCache:2023:Lepers,BonsaiKV:2023:Cai}. 
Orthogonally, \SYS uses \pmem as part of a persistent storage hierarchy.}

\section{Conclusion}
\label{sec:conclusion}

We presented \SYS, a novel storage middleware that accelerates the performance of LSM KVS using a heterogeneous storage hierarchy.
Contrary to prior work, \SYS consolidates the inherent properties of LSM with the parallelism, I/O bandwidth, and capacity of different storage devices. 
Our extensive evaluation shows significant performance improvements over general-purpose systems and specialized KVS with native support for heterogeneous storage. 
\begin{acks}
\revision{This work was conducted when Zhongjie Wu was a student at McGill University.
We thank our shepherd and the anonymous reviewers for their insightful comments and feedback.
We also thank Cláudia Brito, Tânia Esteves, and Willy Zwaenepoel for their input in previous versions of this work.
This work was partially supported by: 
the NSERC Discovery Grant RGPIN-2021-02662; 
the European Regional Development Fund through the NORTE 2030 Regional Programme under Portugal 2030, within project BCDSM (ref.14436, NORTE2030-FEDER-00584600) (João Paulo); 
the Component 5 - Capitalization and Business Innovation, integrated in the Resilience Dimension of the Recovery and Resilience Plan within the scope of the Recovery and Resilience Mechanism of the European Union, framed in the Next Generation EU, for the period 2021-2026, within project ATE (ref.56) (Rúben Adão); 
and by the Portuguese funding agency -- Fundação para a Ciência e Tecnologia (FCT) -- within the PhD grant 2024.02442.BD and project LA/P/0063/2020 (DOI 10.54499/LA/P/0063/2020) (Rúben Adão).}
\end{acks}

\balance

\bibliographystyle{ACM-Reference-Format}
\bibliography{bibliography}

\end{document}